\documentclass[journal]{IEEEtran}
\usepackage[utf8]{inputenc}
\usepackage[T1]{fontenc}
\usepackage{lmodern}
\usepackage{amsmath,amssymb,amsfonts}
\usepackage{graphicx}
\usepackage{subcaption}
\usepackage{float}
\usepackage{url}
\usepackage{cite}
\usepackage{geometry}
\usepackage{csquotes}
\title{\textbf{The Unspoken Crisis of Learning: The Surging Zone of No Development}}

\author{
Euzeli C. dos Santos Jr., and Tracey Birdwell\\
\textit{Purdue University, Indianapolis, IN, USA}\\
\texttt{edossant@purdue.edu}
}

\date{Manuscript submitted October 2025}

\begin{document}
\maketitle

\begin{abstract}
AI has redefined the boundaries of assistance in education, often blurring the line between guided learning and dependency. This paper revisits Vygotsky’s Zone of Proximal Development (ZPD) through the lens of the P2P Teaching framework. By contrasting temporary scaffolding with the emerging phenomenon of permanent digital mediation, the study introduces the concept of the Zone of No Development (ZND), a state in which continuous assistance replaces cognitive struggle and impedes intellectual autonomy. Through theoretical synthesis and framework design, P2P Teaching demonstrates how deliberate disconnection and ethical fading can restore the learner’s agency, ensuring that technological tools enhance rather than replace developmental effort. The paper argues that productive struggle, self-regulation, and first-principles reasoning remain essential for durable learning, and that responsible use of AI in education must include explicit mechanisms to end its help when mastery begins. 
\end{abstract}


\section{Introduction}

\begin{figure}[!b]
\centering
\begin{subfigure}[b]{0.8\linewidth}
    \centering
    \includegraphics[width=\linewidth]{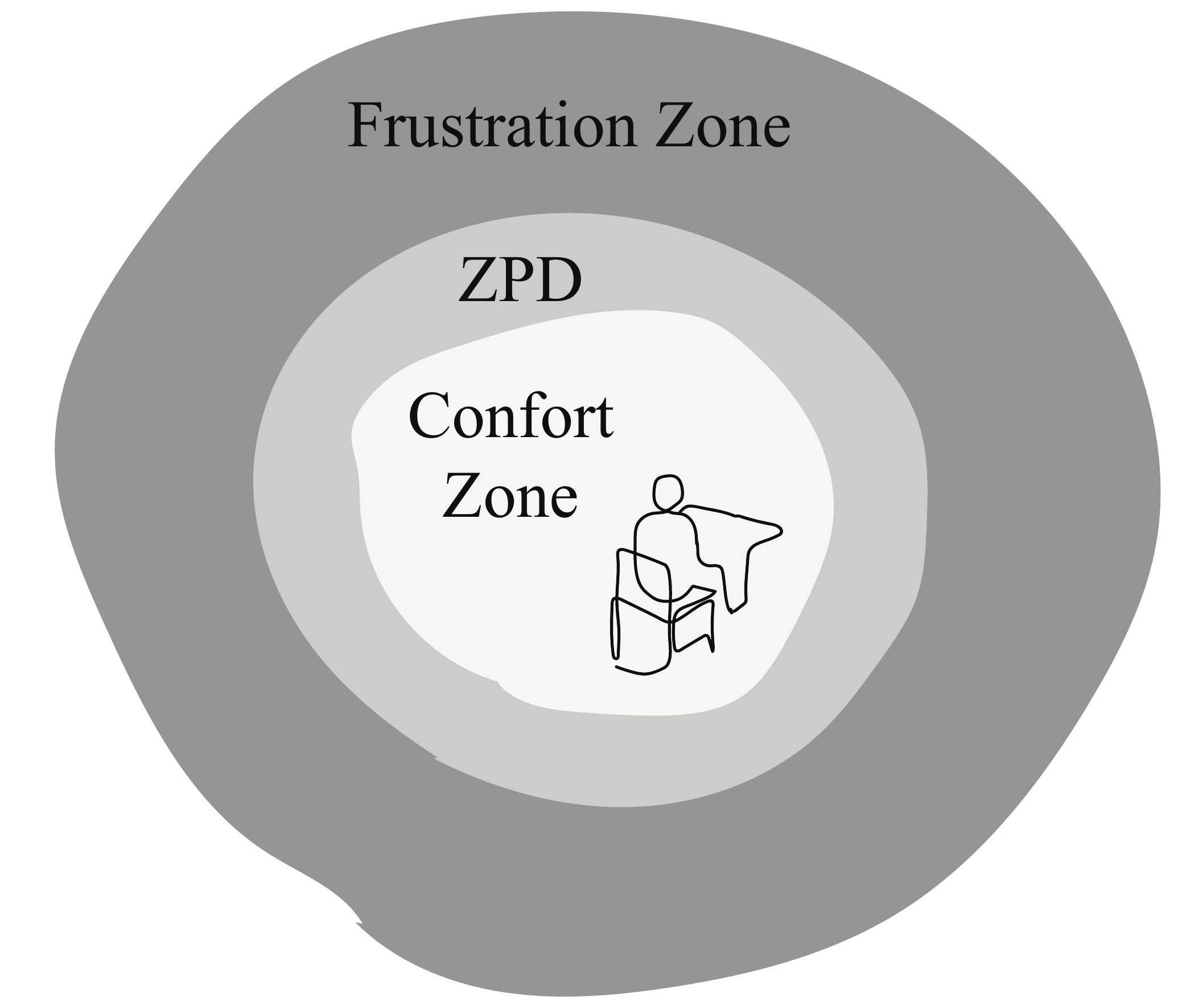}
    \caption{}
\end{subfigure}
\begin{subfigure}[b]{0.8\linewidth}
    \centering
    \includegraphics[width=\linewidth]{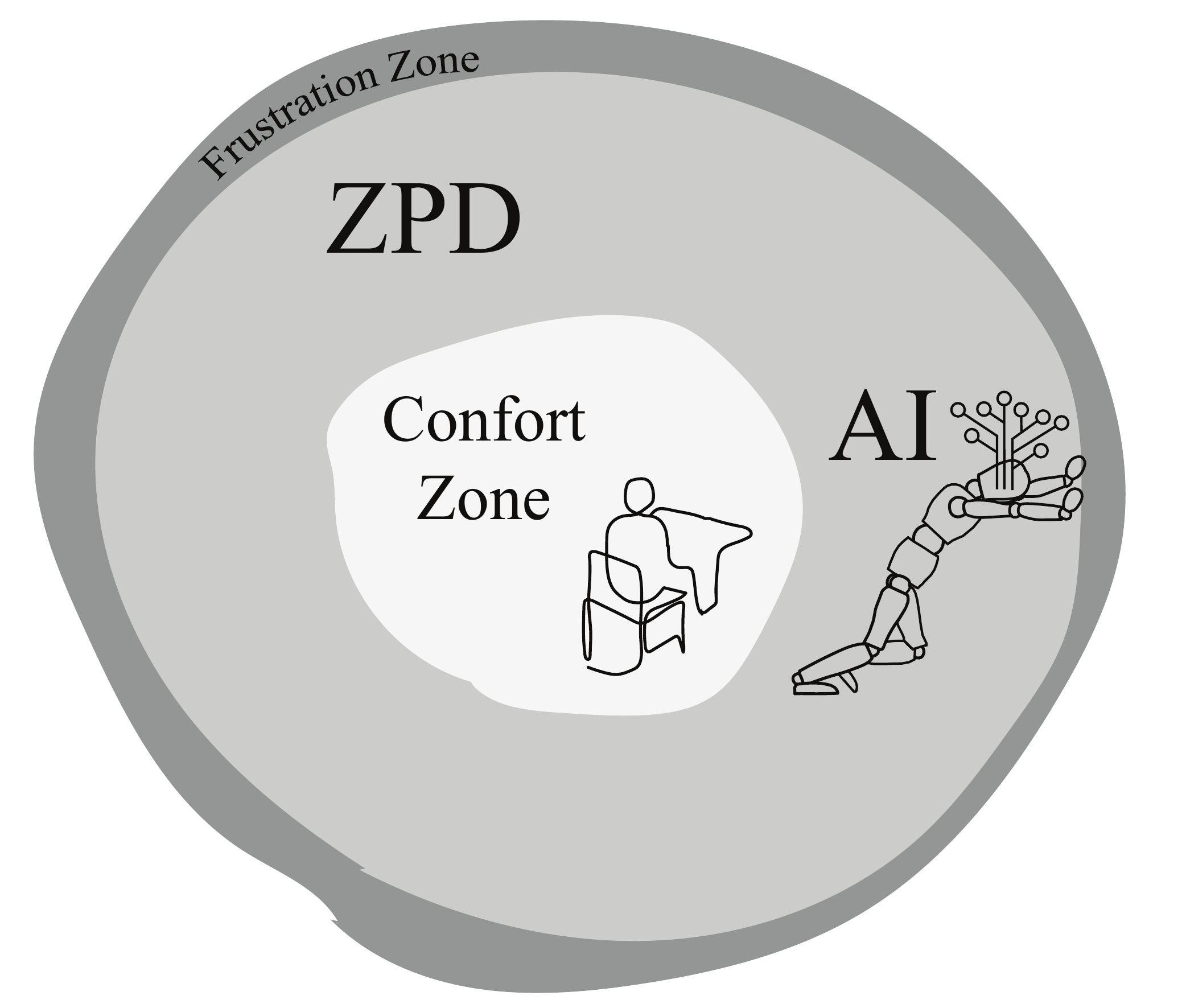}
    \caption{}
\end{subfigure}
\caption{(a) Classical model showing the balance between the comfort, ZPD, and frustration zones. (b) AI-assisted model where continuous support reduces the frustration zone and risks shrinking genuine ZPD.}

\label{fig:ZPD}
\end{figure}

The ZPD originates in Vygotskian sociocultural theory as the distance between what a learner can accomplish independently and what becomes achievable with guidance from a more knowledgeable other \cite{vygotsky1978,vygotsky1986}. As a foundational construct in the learning sciences, ZPD reframed cognitive development as fundamentally mediated by social interaction, language, and tools. Subsequent scholarship connected ZPD to the notion of instructional scaffolding, graduated support that is purposefully introduced and then faded as competence grows, articulated classically by Wood, Bruner, and Ross \cite{wood1976} and elaborated in work on apprenticeship and guided participation \cite{rogoff1990}. Together, these contributions established ZPD as a central analytic and design lens for teaching, aligning curriculum, assessment, and pedagogy with learners’ evolving capabilities and the mediational means that enable them. 

Fig. \ref{fig:ZPD}(a) illustrates the foundational structure of Vygotsky’s concept of the Zone of Proximal Development (ZPD). The innermost region, labeled \textit{Comfort Zone}, represents what the learner can accomplish independently with existing knowledge and skills. While safe and familiar, learning that remains confined to this region leads to minimal cognitive growth. The surrounding layer, i.e., the \textit{Zone of Proximal Development - ZPD}, defines the space where meaningful learning occurs: tasks that a learner cannot yet complete alone but can accomplish successfully with guidance from a more knowledgeable other. Instruction within the ZPD relies on a process in which support is gradually reduced as the learner gains mastery, i.e., scaffolding \cite{wood1976,rogoff1990}. This dynamic interaction between autonomy and guided participation represents the optimal condition for conceptual change and skill development. The outermost region of the figure, the \textit{Frustration Zone}, encompasses tasks that exceed the learner’s current capability even with assistance. Operating predominantly within this zone often results in cognitive overload, confusion, and disengagement. Contemporary research in STEM and engineering education confirms that learning effectiveness is maximized when instruction keeps students within or near their ZPD through adaptive scaffolding and collaborative problem-solving \cite{hmelo2007,princefelder2006}. 

Indeed, in STEM classrooms, guided inquiry and problem-based learning leverage ZPD by calibrating supports to domain practices and gradually transferring responsibility to students \cite{hmelo2007}. Within engineering education, evidence for active and inductive teaching approaches shows that structured guidance, i.e., concept questions, worked-example progressions, and project scaffolds, improves conceptual gains and problem-solving, consistent with working in learners’ ZPD \cite{prince2004,princefelder2006}. Collectively, these strands position ZPD as a practical, theory-grounded compass for designing AI-mediated and instructor-mediated supports that cultivate disciplinary thinking in STEM and engineering.

\section{The Concept of Zone of No Development}

Within the Zone of Proximal Development (ZPD) framework, as previously discussed, the presence of a \textit{more knowledgeable other} is fundamental for facilitating cognitive growth. Equally essential, however, is the gradual withdrawal of this external support, a process known as \textit{scaffolding fade-out}, which enables the learner’s \textit{comfort zone} to expand through autonomous mastery. In traditional instructional settings, this fading process occurs naturally as the instructor recognizes increasing learner competence. In contrast, when artificial intelligence (AI) becomes the primary mediating agent, the persistence of constant and immediate assistance can alter this dynamic. Rather than promoting independence, AI’s continuous availability may unintentionally inhibit the natural reduction of guidance that fosters deeper learning. Initially, this ever-present support appears beneficial, as it effectively reduces the \textit{frustration zone} by keeping the learner within a seemingly enlarged ZPD, see Fig. \ref{fig:ZPD}(b). However, such expansion can be pedagogically deceptive: while it gives the impression of improved accessibility and understanding, it risks diminishing the learner’s cognitive resilience and capacity for self-regulated reasoning over time.

\begin{figure}[!b]
\centering
\begin{subfigure}[b]{0.9\linewidth}
    \centering
    \includegraphics[width=\linewidth]{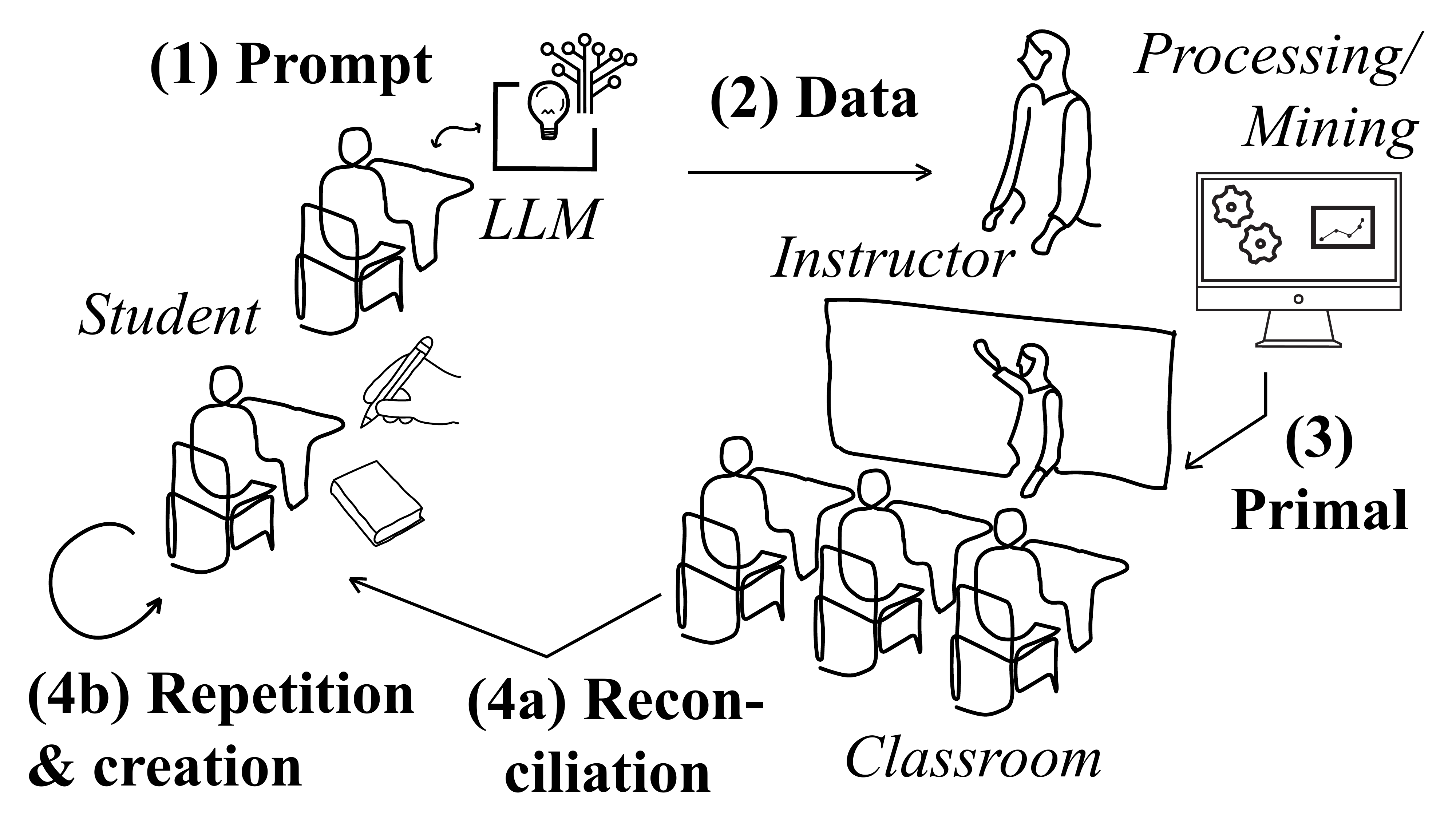}
    \caption{}
\end{subfigure}
\begin{subfigure}[b]{0.9\linewidth}
    \centering
    \includegraphics[width=\linewidth]{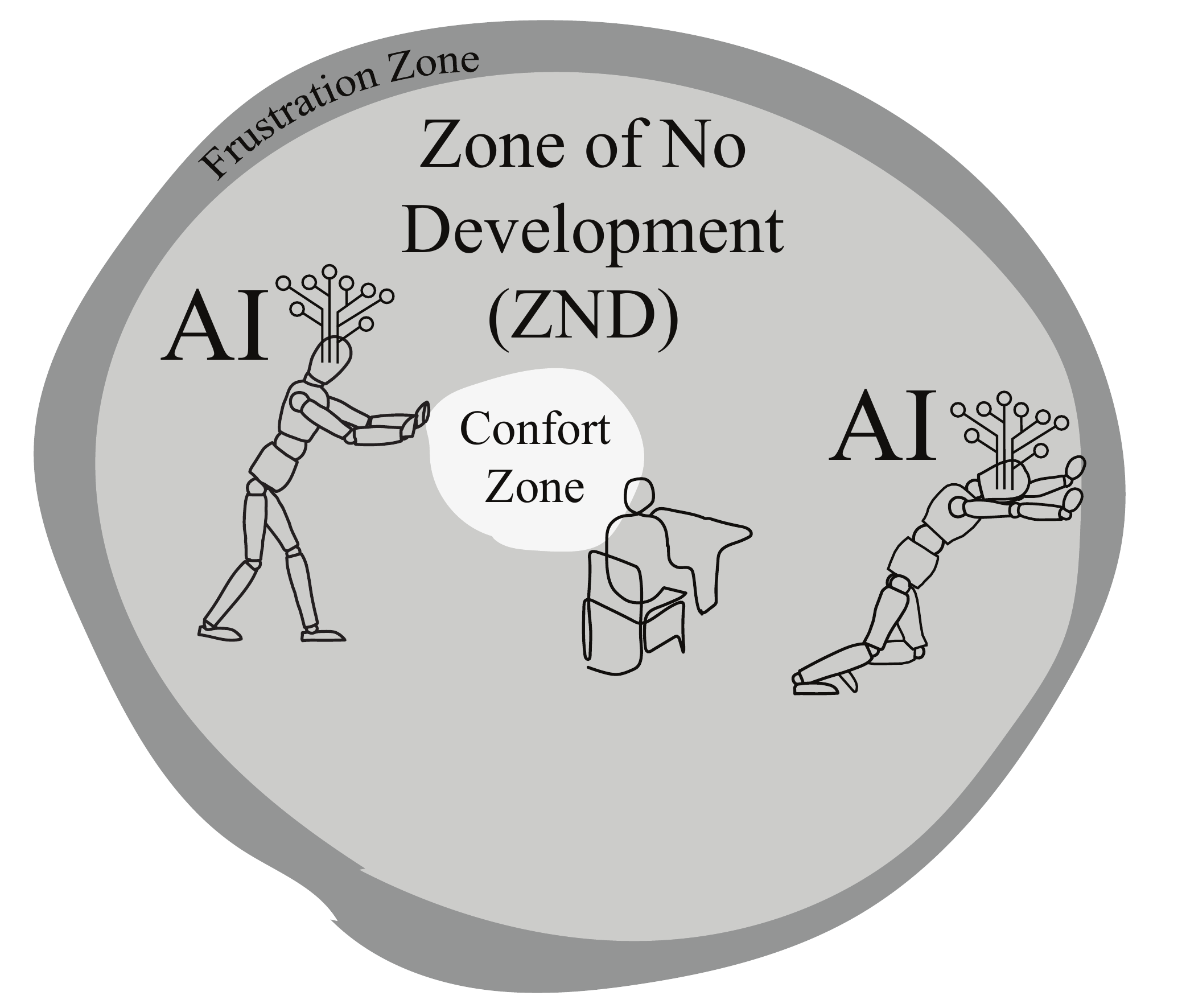}
    \caption{}
\end{subfigure}
\caption{(a) P2P teaching framework. (b) P2P teaching's Phase 4b (left) and the impact of AI shrinking ZPD (right).}
\label{fig:P2P_ZPD}
\end{figure}

Fig. \ref{fig:P2P_ZPD}(a) shows the Prompt-to-Primal (P2P) teaching framework proposed in \cite{dosSantos2025}. This is an AI-integrated pedagogical approach that begins with student-generated prompts and culminates in instructor-guided reasoning grounded in first principles, transforming AI interaction into a catalyst for deep, self-regulated learning.

In this context, the rigid instructional frame proposed by the Prompt-to-Primal (P2P) Teaching framework plays a crucial role in ensuring that students engage with AI support responsibly and developmentally. By defining structured phases, i.e., from prompt generation to primal reasoning and reflective creation, P2P Teaching preserves the essence of the Zone of Proximal Development (ZPD) while preventing the risks associated with perpetual assistance since Phase 4b restric the use of LLM and require the students to use the traditional way of learning (pencil and paper). As illustrated in Fig.~\ref{fig:ZPD}(b), when AI continuously mediates learning, it may appear to expand the learner’s ZPD by reducing the frustration zone. Yet, as the cognitive struggle diminishes, so does the learner’s agency in expanding their comfort zone, ultimately threatening the transition from supported to autonomous reasoning. Within the P2P framework, structured constraints, deliberate scaffolding fade-out, and guided reflection counteract this effect, ensuring that learners remain intellectually challenged and self-regulated. Without such design, the ZPD risks collapsing into what can be described as a \textit{Zone of No Development} (ZND), i.e., a state in which comfort and dependency prevail over curiosity and cognitive growth.

The emergence of what can be called the illusion of learning is an inevitable byproduct of the constant presence of artificial intelligence within the learner’s cognitive environment. This illusion is rooted in the discrepancy between perceived progress and genuine cognitive development, i.e., a phenomenon well documented in educational psychology through the concepts of fluency illusion and overconfidence bias. When AI systems provide instantaneous feedback, perfectly phrased explanations, or complete problem solutions, students may misinterpret ease of access and speed of comprehension as indicators of mastery. However, mastery that is not accompanied by retrieval effort, reflection, and error correction lacks the neural consolidation necessary for durable learning. Within the Zone of Proximal Development (ZPD), authentic growth occurs only when learners actively reconstruct understanding, struggle productively, and eventually internalize strategies that were once scaffolded by a more knowledgeable other. When this developmental tension is removed (because AI’s assistance never fades) the learner’s comfort zone ceases to expand, and the ZPD can gradually transform into a Zone of No Development (ZND). In this state, cognitive comfort replaces epistemic curiosity; performance replaces understanding; and the illusion of learning becomes both symptom and cause of stagnation. The P2P Teaching framework explicitly addresses this risk by embedding reflective and primal stages that compel the learner to re-engage with first principles. More importantly, P2P Teaching explicitly requires learners to disconnect from AI tools during the Creation phase (Phase 4b). In doing so, it restores the necessary friction between comfort and frustration, an equilibrium essential to sustain authentic learning and prevent the descent from ZPD to ZND.

\section{Temporary versus Permanent Scaffolding}

The concept of scaffolding, as originally derived from Vygotsky’s Zone of Proximal Development (ZPD), refers to temporary instructional support that assists learners in completing tasks slightly beyond their independent capability. Such support is gradually withdrawn as the learner internalizes the processes and develops autonomy. This notion of temporary scaffolding is central to sustainable learning because it preserves the dynamic balance between assistance and independence, allowing the comfort zone to expand through genuine cognitive effort. In contrast, the introduction of persistent or algorithmic guidance, such as that offered by large language models (LLMs), risks transforming scaffolding into a permanent condition, thereby distorting the natural rhythm of cognitive development.

\begin{figure*}[!t]
\centering
\includegraphics[width=0.8\linewidth]{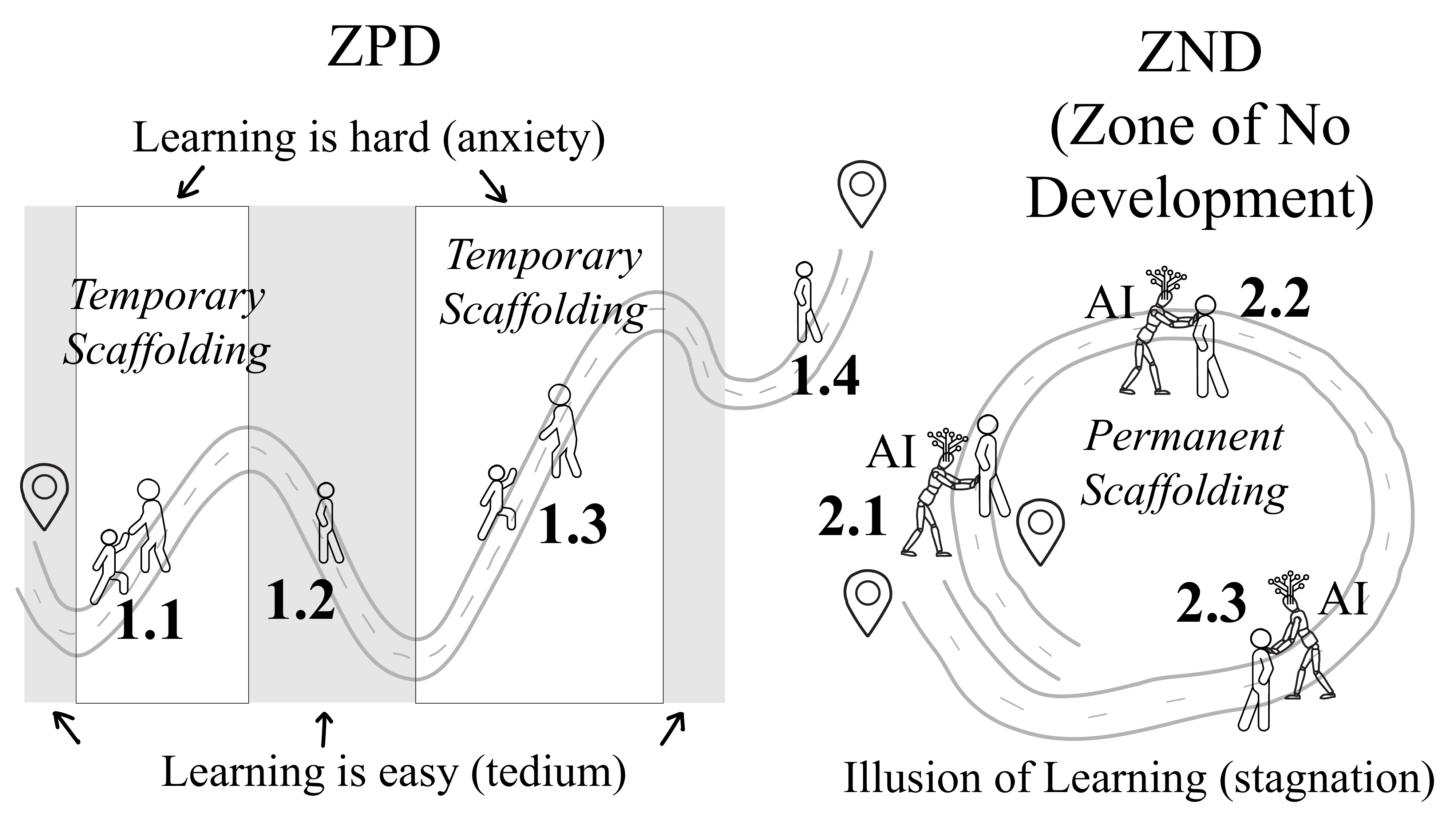}
\caption{Comparative representation of temporary versus permanent scaffolding: the left trajectory (1.1)–(1.4) depicts traditional learning progression within ZPD: (1.1) learning with guided assistance, (1.2) learning without assistance, (1.3) reduced guidance during learning, and (1.4) independent learning—illustrating the natural fade-out of scaffolding and the expansion of the learner’s comfort zone; the right trajectory (2.1)–(2.3) represents learning under continuous AI mediation: (2.1) initial AI-assisted engagement, (2.2) dependency formation due to persistent assistance, and (2.3) the illusion of learning leading to the Zone of No Development (ZND).}
\label{fig:ZPD_ZND}
\end{figure*}

Fig.~\ref{fig:ZPD_ZND} illustrates this contrast through two independent concepts. The left sequence (1.1–1.4) depicts the traditional model of guided learning. In (1.1), the learner operates within the ZPD under structured guidance, this is the stage of learning with guided assistance, in which scaffolding from a more knowledgeable other is fully present. In (1.2), learning without assistance begins to emerge as temporary supports are deliberately faded, enabling the learner to practice and consolidate knowledge independently. By (1.3), the learner experiences less assistance while learning, meaning that guidance is available only when conceptual breakdowns occur, reinforcing metacognitive regulation and self-efficacy. Finally, (1.4) represents independent learning, the culmination of successful scaffolding fade-out. Here, cognitive load is appropriately balanced: learning is neither too easy (leading to tedium) nor too difficult (leading to anxiety), but remains productive and adaptive. This sequence models the core principle that effective scaffolding must be temporary, intentionally removed once competence is achieved.

The right-side of the figure, i.e., sequences from (2.1) to (2.3) in Fig.~\ref{fig:ZPD_ZND} represents learning under continuous AI assistance, which represents permanent scaffolding. In (2.1), learners engage with AI-mediated explanations and immediate feedback, which can initially mirror the benefits of human tutoring by maintaining engagement and reducing frustration. However, as shown in (2.2), persistent availability of assistance erodes the learner’s need to retrieve, struggle, and reconstruct understanding, leading to dependency and reduced cognitive autonomy. By (2.3), the learner enters a state of illusion of learning, characterized by high fluency but low retention, a phenomenon empirically associated with overconfidence bias and performance, i.e., learning dissociation \cite{bjork2013, soderstrom2015}. The ZPD gradually collapses into what is described as the Zone of No Development (ZND), where perceived mastery replaces genuine understanding. 

The left trajectory in Fig.~\ref{fig:ZPD_ZND} therefore corresponds to temporary scaffolding, where supports are dynamic and finite, while the right trajectory illustrates permanent scaffolding, where assistance becomes static and omnipresent. The pedagogical risk of permanent scaffolding is not the presence of AI itself but its persistence. Without designed constraints, such as those embedded in the Prompt-to-Primal (P2P) Teaching framework, AI can stabilize learners in a comfortable but stagnant cognitive space. P2P Teaching mitigates this risk by enforcing structured disconnection during the Creation Phase (Phase 4b), ensuring that students experience the productive tension necessary for true internalization. 

\section{When Help Becomes Harm – The Ethics of Fading}

The ethical dimension of teaching has always extended beyond content mastery, encompassing the formation of learners capable of independent and responsible thought. Within the context of the Prompt-to-Primal (P2P) framework, this ethical dimension reemerges through the notion of \textit{fading as a moral act}. When scaffolding persists indefinitely—whether in the form of constant AI availability or unrestricted electronic access—the educator risks transforming guidance into dependence. In contrast, ethical fading deliberately restores the learner’s autonomy by re-establishing productive struggle as a natural and necessary stage of intellectual growth. In P2P Teaching, the instructor’s responsibility is not only to provide access to knowledge but also to withdraw assistance at the precise moment when dependency threatens development. This act, while pedagogical in method, is ethical in nature: it affirms that genuine learning requires effort, delay, and cognitive resistance, not merely fluency or ease of retrieval.

An essential component of this ethical stance involves designing strategies that convince students to \textit{put electronics away} (literally and metaphorically) to rediscover the discipline of uninterrupted thinking. Such moments of technological silence are not nostalgic gestures but epistemological necessities. They invite students to re-engage with the discomfort of uncertainty, the cognitive friction of problem-solving, and the satisfaction of self-derived insight. Within the P2P framework, these practices operationalize the “primal” stage of learning, in which knowledge is reconstructed from first principles, without the mediation of an ever-present digital tutor. Ethical teaching, therefore, demands intentional creation of environments where struggle is not avoided but welcomed as the authentic space of the Zone of Proximal Development (ZPD).

Assessment practices must also embody this principle. Examinations designed with no open books, no internet, and no AI assistance reaffirm the central message that what is being evaluated is not the student’s capacity to prompt the right question but to internalize and retrieve knowledge independently. Such conditions might appear restrictive in an era that celebrates access and connectivity, yet they restore balance to an educational ecosystem that risks collapsing into perpetual mediation. In these “closed-world” settings, the student’s reasoning is tested as an internal process, not as an externally scaffolded exchange—a distinction vital for sustaining the integrity of the ZPD and preventing its degradation into a Zone of No Development (ZND). 

Finally, the ethics of fading extends to the implicit moral contract between instructor and learner. Even in the absence of proctoring or surveillance, P2P Teaching presupposes an explicit statement of expectation: students are trusted, and required, to disengage from AI during certain phases of learning. What once seemed self-evident to earlier generations (the act of studying without technological mediation) must now be re-articulated as an ethical norm. This expectation is not punitive but developmental; it cultivates intellectual honesty and the capacity for self-regulation, qualities essential for lifelong learning. In this sense, fading is not only a pedagogical device but also a declaration of respect for the learner’s potential. To withhold help, at the right time and for the right reason, is to affirm the belief that the student can (and must) think independently.

\section{Learning $versus$ Intellectual Autonomy}

Learning, as treated in the paper, refers to the process of constructing durable knowledge, i.e., knowledge that is retrievable, transferable, and flexible. It requires cognitive effort, productive struggle, and the gradual internalization of strategies once scaffolded by a more knowledgeable other.

Intellectual autonomy, on the other hand, refers to the learner’s ability to reason, problem-solve, and self-regulate without external mediation. Autonomy emerges from learning, but it is not identical to learning. One can momentarily “perform” (especially with access to AI tools) without having developed autonomy, this is the core mechanism of the illusion of learning described as observed in Fig. \ref{fig:ZPD_ZND}(b). 

The argument is that continuous AI assistance blurs the boundary between performance and autonomy, enabling students to complete tasks but preventing the development of the independence required to extend, adapt, or creatively apply what they know. In that sense, autonomy is treated as a long-term developmental outcome of learning, not merely as the immediate ability to complete a task.

\section{Conclusion}

The intersection between Vygotsky’s Zone of Proximal Development (ZPD) and AI-mediated learning reveals both promise and peril. While intelligent systems can extend access to guidance and information, they can also undermine the very conditions that make learning transformative. This paper has hypothesized that when scaffolding becomes permanent, i.e., sustained by continuous AI assistance, the learner’s capacity for autonomous reasoning diminishes, leading to what is here defined as the Zone of No Development (ZND). Within this state, comfort replaces challenge and fluency replaces understanding, resulting in an illusion of learning that obstructs intellectual growth.

The Prompt-to-Primal (P2P) Teaching framework offers a pathway to mitigate this risk by embedding structured phases of connection, disconnection, and reflection. Its deliberate fading mechanisms, particularly in the Creation phase, ensure that learners regain agency and cultivate cognitive endurance. Ethically, this design reaffirms that authentic education involves struggle, delay, and effort, i.e., conditions that technology should amplify but never abolish.

Ultimately, the responsible integration of AI in education depends on the understanding to know when to step away from it. Only by reinstating the necessity of productive struggle and the ethics of fading can educators preserve the integrity of the ZPD and prevent its descent into the Zone of No Development. 

\bibliographystyle{IEEEtran}

\end{document}